\documentclass{emulateapj}

\usepackage{color}

\shorttitle{Nonthermal X-ray Emission from Cassiopeia A}
\shortauthors{Patnaude \& Fesen}

\begin{document}

\title{Proper Motions and Brightness Variations of Nonthermal X-ray
Filaments in the Cassiopeia A Supernova Remnant}

\author{Daniel J. Patnaude\altaffilmark{1} \& Robert A. Fesen\altaffilmark{2}}

\altaffiltext{1}{Smithsonian Astrophysical Observatory, Cambridge, MA 02138}

\altaffiltext{2}{Department of Physics and Astronomy, 6127 Wilder Lab, 
Dartmouth College, Hanover, NH 03755}

\begin{abstract}

We present {\it Chandra} ACIS X-ray observations of the Galactic supernova
remnant Cassiopeia A taken in December 2007. Combining these data with previous
archival {\it Chandra} observations taken in 2000, 2002, and 2004, we estimate
the remnant's forward shock velocity at various points around the outermost
shell to range between 4200 and 5200 $\pm500 $ km s$^{-1}$.  Using these
results together with previous analyses of Cas A's X-ray emission, we present a
model for the evolution of Cas A and find that it's expansion is well fit by a
$\rho_{ej} \propto r^{-(7-9)}$ ejecta profile running into a circumstellar 
wind.
We further find that while the position of the reverse shock in this model is
consistent with that measured in the X-rays, in order to match the forward
shock velocity and radius we had to assume that $\sim$ 30\% of the explosion
energy has gone into accelerating cosmic rays at the forward shock.  The new
X-ray images also show that brightness variations can occur for some forward
shock filaments like that seen for several nonthermal filaments seen projected
in the interior of the remnant. Spectral fits to exterior forward shock
filaments and interior nonthermal filaments show that they exhibit similar
spectra. This together with similar flux variations suggests that interior
nonthermal filaments might be simply forward shock filaments seen in
projection and not located at the reverse shock as has been recently proposed.

\end{abstract}


\section{Introduction}

Cassiopeia A (Cas A) is one of the youngest known Galactic supernova remnants
(SNR) with an estimated explosion date no earlier than $1681 \pm19$
\citep{fesen06}. Optical echoes of the supernova outburst have been recently
detected \citep{Rest08}, the spectra of which indicate Cas~A is the remnant of
a Type IIb supernova event \citep{Krause08} probably from a red supergiant in
the mass range of 15--25 M$_{\odot}$ that may have lost much its hydrogen envelope
to a binary interaction \citep{Young06}.

Viewed in X-rays, the remnant consists of a line emitting shell arising from
reverse shocked ejecta rich in O, Si, Ar, Ca, and Fe
\citep{fabian80,markert83,vink96,hughes00,will02,will03,hwang03,laming03}.
Exterior to this shell are faint X-ray filaments which mark the current
position of the remnant's forward shock front. The emission found here is
nonthermal X-ray synchrotron radiation as well as faint line emission from
shocked circumstellar material (CSM).  

\citet{vink98} compared {\sl Einstein} HRI to {\sl ROSAT} HRI observations of
Cas A to measure the expansion of the bright shell, finding an expansion age of
$\sim$ 500 yr, considerably less than the $\sim$ 800 yr expansion age derived
from 1.5 and 5.0 GHz radio observations \citep{anderson95}, but similar to the
400--500 yr expansion age found by \citet{agueros99} using data taken at 151 MHz.  More
recently, \citet{delaney03} using {\sl Chandra} X-ray observations taken in 2000 and
2002 presented the first proper motion measurements of the forward blastwave
velocity. Assuming a distance of 3.4 kpc \citep{reed95}, they
estimated a blast wave expansion velocity of $\approx$ 5000 km s$^{-1}$. 

Besides the outlying nonthermal emission filaments associated with the
forward shock, some filamentary nonthermal X-ray emission is also seen in
projection in the interior of the SNR \citep{delaney04a}. Whether these
interior filamentary emissions originate from a wrinkled forward shock seen in
projection or arises from nonthermal emission mechanisms in the interior of the
SNR is currently uncertain \citep{laming01,uchiyama08,helder08}. 

Comparisons of {\it Chandra} observations taken in 2000, 2002, and 2004
revealed secular changes in several X-ray thermal knots and in one nonthermal
filament projected in the remnant's interior \citep{patnaude07}.
\citet{uchiyama08} using the same multi-epoch {\it Chandra} observations 
found evidence for rapid variability in many more interior nonthermal X-ray emission
filaments. Motivated by similar
changes seen in RX J1713-3946 \citep{uchiyama07}, they measured the time
variability of selected filaments to determine the local magnetic field
strength in the variable regions.  Their results suggest that the magnetic
field in these regions is relatively high, B $\sim$ 1 mG. 

Such a high magnetic field strength would be consistent with equipartition field
strengths inferred in observations of bright radio knots in the remnant
\citep{longair94,wright99}.  \citet{uchiyama08} argue that their result points
to a synchrotron origin for the emission coming from these knots, ruling out
nonthermal bremsstrahlung from $\sim$ 100 keV electrons \citep{laming01}, and
suggest that this is strong evidence for a hadronic origin to the TeV emission
observed in Cas A \citep{aharonian01,albert07}. Based on the location of the
synchrotron knots, Uchiyama \& Aharionian suggest that the emission is
located primarily at the reverse shock, and \citet{helder08} reach a similar
conclusion.

Here we present forward shock velocity measurements using new {\it Chandra}
ACIS observations of Cas A taken in December 2007 and compare these results to
models for SNR evolution with and without efficient shock acceleration.  The
new observations show that many nonthermal emission filaments and features have
undergone substantial brightness variations over the last four years.  Model
fits to the nonthermal emission coming from both the forward shock and the
interior filaments indicate that they are quantitatively similar.  We also
present evidence for fast variability in forward shock front filaments which
argues against the conclusion that rapid variability is a property restricted
to emission at the reverse shock.

\section{Observations}

\begin{figure}
\plotone{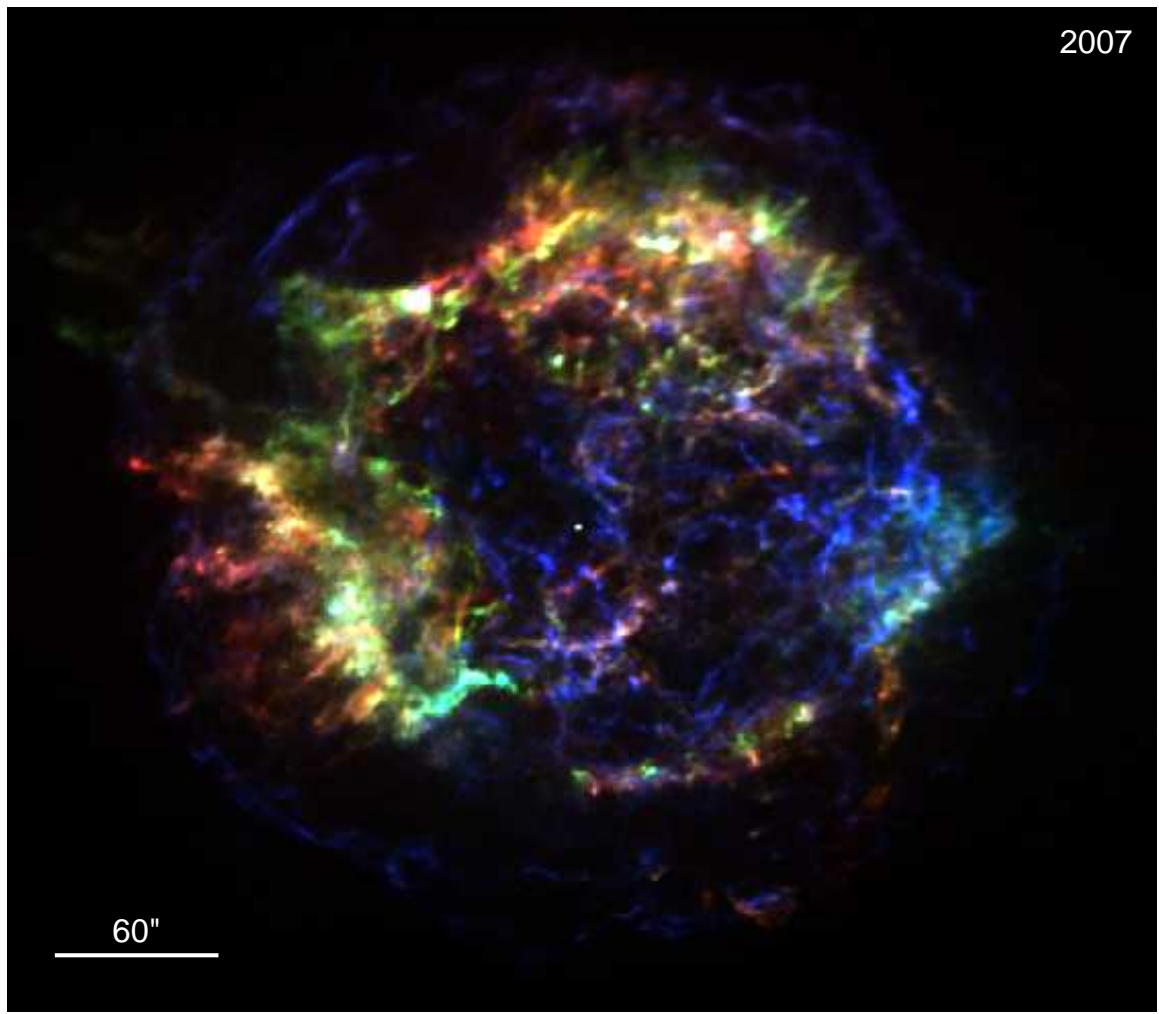}
\caption{Exposure corrected RGB color December 2007 image of Cas A. 
Red corresponds to 
0.5--1.5 keV, green to 1.5--3.0 keV, and blue to 4.0--6.0 keV.
This figure is available as part of an on-line animation
in the electronic edition of the {\it Astrophysical Journal}, which
shows the dynamical and spectral evolution of Cassiopeia A from Jan 2000 to Dec 2007.}
\label{fig:casa07}
\end{figure}

Cas A was observed with the ACIS-S3 chip on {\sl Chandra} in two 25 ksec
observations taken on 5 Dec 2007 (ObsID 09117) and 8 Dec 2007 (ObsID 09773).
The ACIS's $0\farcs492$ CCD pixel scale under-samples the telescope's $\simeq 0
\farcs5$ resolution.  The data were reprocessed using CIAO 4.0.1 and the latest
version of the {\it Chandra} CalDB (Version 3.4.2).  Figure~\ref{fig:casa07}
shows the combined, exposure corrected image coded by energy.  Red corresponds
to 0.5--1.5 keV, green to 1.5--3.0 keV, and blue to 4.0--6.0 keV.  

For our analyses, we also made use of previous {\sl Chandra} ACIS
observations taken on 30 Jan 2000 (ObsID 00114; PI: Holt), 6 Feb 2002 (ObsID
01952; PI: Rudnick), and 8 Feb 2004 (ObsID 05196; PI: Hwang).  These archival
data were also reprocessed using the latest version of the CalDB and all four
ACIS images were projected to a common tangent point, chosen to be the
expansion center determined by \citet{thorstensen01}. Finally, the
images were registered against the central compact object (CCO). 
Unregistered, the centroid of the Cas A CCO differs by $0 \farcs 08$ between
2000 and 2002, and by $0 \farcs 33$ between 2000 and 2008. We have registered
the images against the year 2000 observations, though we note that when 
performing the same analysis on the unregistered images, we found no significant 
differences in our results.

To avoid the problems with bad columns and node boundaries discussed by
\citet{delaney03}, exposure corrected images for the 2000 and 2007 observations
were created assuming a 1.85 keV source.  We note that using a mono-energetic
correction results in an artificially higher surface brightness for the forward
shock filaments. 

\section{Results and Analysis}

\subsection{Proper Motion of the Forward Shock Front}

Using the ACIS 2000 and 2002 images, \citet{delaney03} estimated the proper
motions of several forward shock, nonthermal filaments around the SNR. Based on
their average estimated proper motion of $0 \farcs 30$ yr$^{-1}$, we expected
the filaments to have shifted by $\simeq 2 \farcs 4$ over 7.87 yr, or
approximately 5 ACIS pixels.  In a follow-up to their work, we used the
locations of the forward shock X-ray filaments on our Dec 2007 ACIS images
compared with their positions on 2000, 2002, and 2004 ACIS images to obtain
improved estimates on the proper motion of the remnant's forward shock front. 

\begin{figure}
\plotone{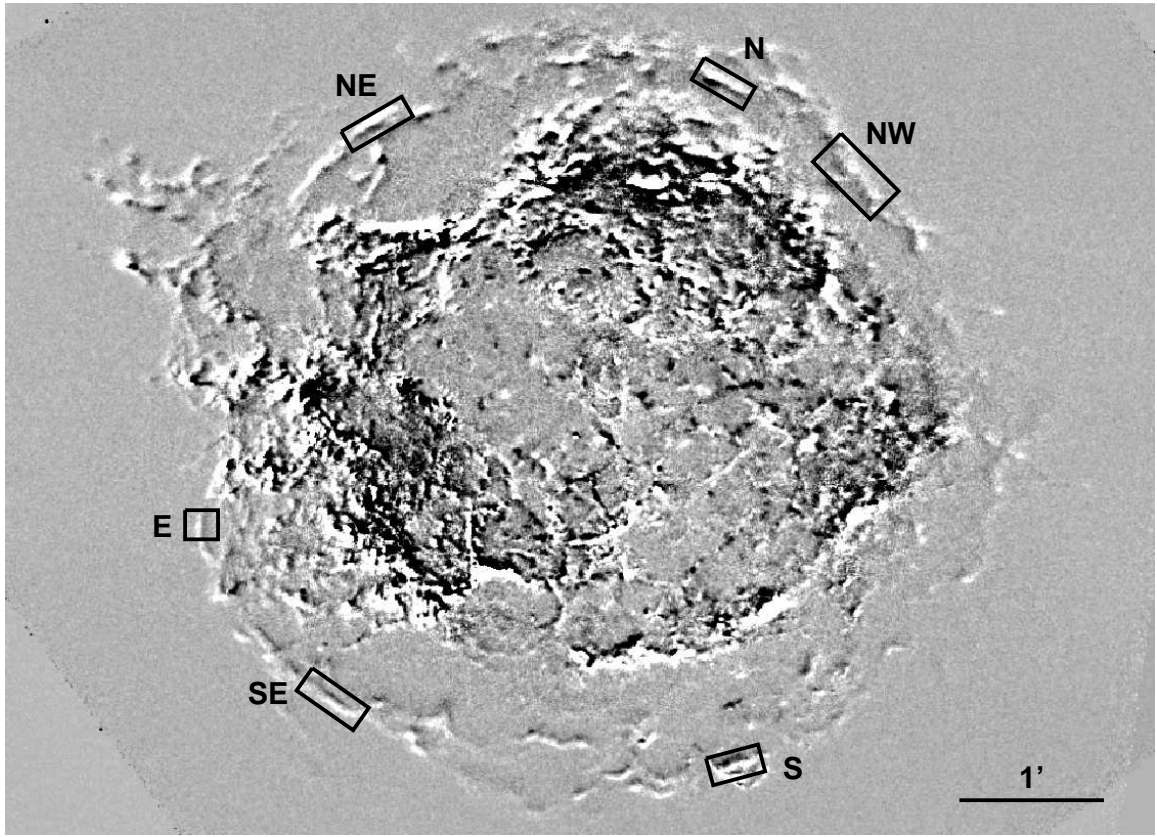}
\caption{A difference image between 2000.08 and 2007.95 {\sl Chandra} 
ACIS images. White correlates
with the direction of filament motion. 
The boxes correspond to regions where we measured
the filament proper motion}
\label{fig:casadiff}
\end{figure}

\begin{figure}
\plotone{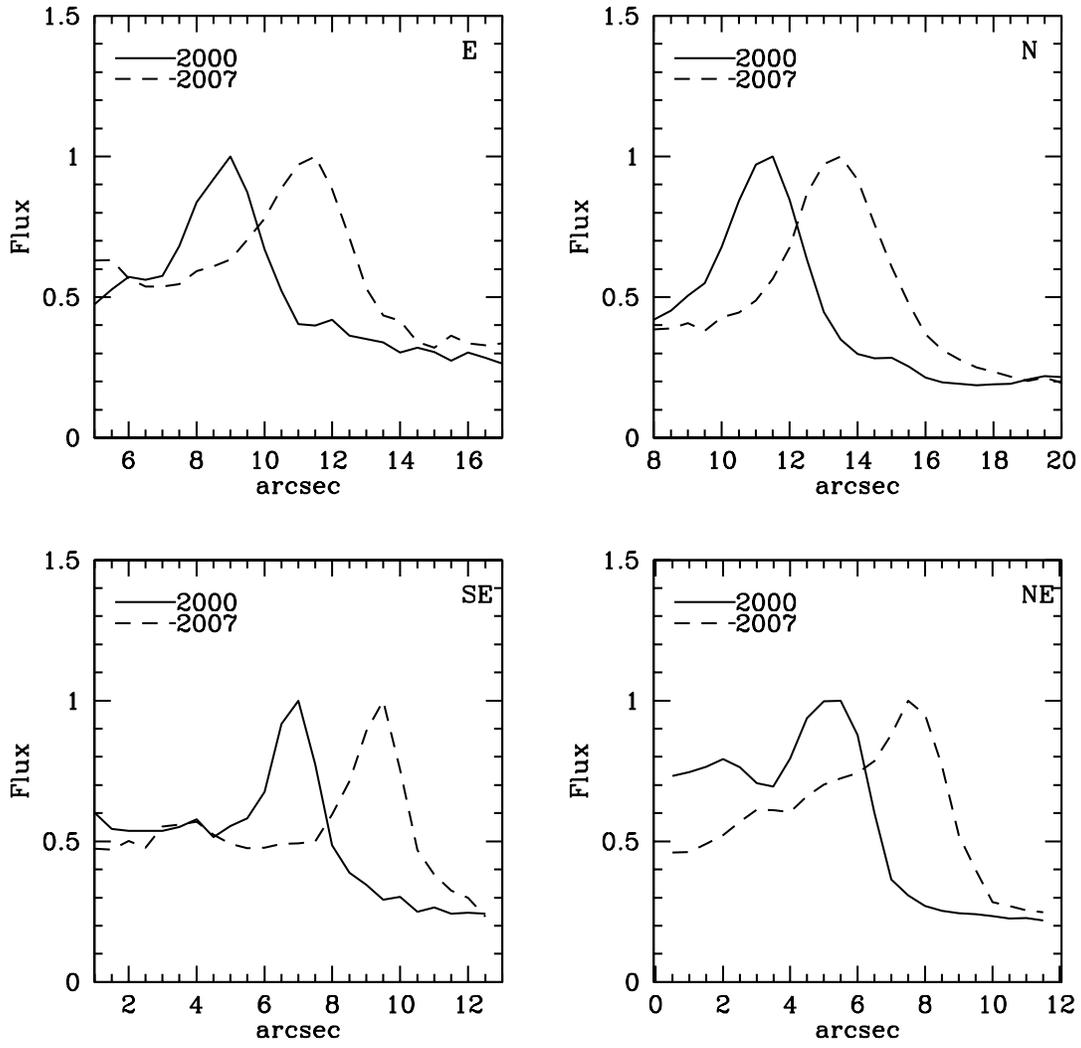}
\caption{Nonthermal, forward shock filament emission profile plots are shown for
four selected regions. Filament profiles
from Jan 2000 and Dec 2007 are shown as solid and dashed lines, respectively. }
\label{fig:profiles}
\end{figure}

Figure~\ref{fig:casadiff} shows a Jan 2000 -- Dec 2007 ACIS difference image of
Cas A.  The six labeled boxes correspond to regions where we measured the
proper motions of the remnant's forward shock filaments.
Figure~\ref{fig:profiles} shows brightness profile plots of four forward shock
filaments taken from the 2000.08 and 2007.95 ACIS images.  As seen in
Figure~\ref{fig:profiles}, there are relatively large and well defined
positional separations between filament positions in the 2000 and 2007 data. 

As noted by \citet{delaney03}, a proper motion measurement using ACIS ideally
should be done using images taken at the same telescope roll angle, as the
telescope point spread function (PSF) varies as a function of azimuthal angle.
Unfortunately, the data taken in 2000 and 2007 are at different roll angles. To
determine the effect that a varying PSF might have on our measurements, we
modeled a 3 keV PSF at each of our chosen positions for the 2000 and 2007
observations. We found that at the average distance of 165$\arcsec$ from the
nominal aimpoint of the observations, the telescope PSF varies by $\leq$ $0
\farcs 05$, much less than an ACIS pixel and well below
the average separations shown in Figure~\ref{fig:profiles}. 

Filament positional shifts were measured two ways. We first fitted a Gaussian
plus background model to the filament profiles and then measured the difference
between the resulting Gaussian centroids. This method is not strictly accurate
because the profiles for nonthermal filaments are not necessarily Gaussian but
are shaped by the swept-up and compressed CSM/ISM magnetic field and vary as a
function of energy \citep{pohl05}.  We also employed a cross-correlation
technique to calculate filament shifts between the two epochs. This technique
has been previously used in measuring proper motions of faint, thin
Balmer-dominated filaments in the Cygnus Loop (see \citealt{patnaude05} for details).

Table~\ref{tab:pm} lists our results for the six selected filament regions
using both measurement techniques.  Using the cross-correlation results, we
estimate proper motions over the nearly eight year time span of 2000.08 to
2007.95 of $0 \farcs 26$ yr$^{-1}$ to  $0 \farcs 32$ yr$^{-1}$ for the six
regions around the SNR, with a typical 1$\sigma$ error of $\pm 0 \farcs 03$
yr$^{-1}$.  

In Table~\ref{tab:pm}, we also list the 2000 -- 2002 proper motion estimates
reported by \citet{delaney03} along with our 2000 -- 2002 measurements but
using our measurement techniques. In general, we find smaller proper motions by
some 15\% -- 20\%.  In view that their quoted errors are comparable or even
smaller than our measurements, we cannot easily account for these differences,
but it may be related to the difference in how their analysis was performed.
Since our results cover nearly four times the time span as their 2000.1 -- 2002.1
proper motion estimates, our results should be more robust.  

\subsection{Cas A's Expansion Velocity and Deceleration}

At a distance of 3.4 kpc, our measured proper range of $0 \farcs 26$ yr$^{-1}$
to  $0 \farcs 32$ yr$^{-1}$ corresponds to forward shock front expansion
velocities of 4200 to 5200 $\pm 500 $ km s$^{-1}$.  The average expansion
velocity for the six regions listed in Table~\ref{tab:pm} is $\approx$ 4900 km
s$^{-1}$, in good agreement with the 5000 km s$^{-1}$ reported by
\citet{delaney03} for some two dozen regions.

\citet{vink98} measured the expansion of Cas A's main shell in X-rays by
comparing ROSAT and Einstein HRI observations that were separated by 17 years.
They found an expansion time-scale of 501 $\pm$ 15 yr, considerably more than
the $\approx$325 yr optically derived age of Cas A
\citep{thorstensen01,fesen06}, but also much less than the reported $\sim$ 800
yr expansion age determined in the radio \citep{anderson95}, based on 
1.5 and 5.0 GHz observations. \citet{agueros99} found an expansion age similar
to \citet{vink98}, from 151 MHz observations.

\citet{gotthelf01} measured the angular size of
Cas A to be 153$\arcsec$ $\pm$ 12$\arcsec$. \citet{thorstensen01} estimate an
undecelerated explosion convergence date of $1671 \pm 1$ based on proper motion
measurements on 17 outlying ejecta knots mainly using archival Palomar 5m
images dating as far back as 1951, while \citet{fesen06} estimated a
convergence date of $1681 \pm 19$ based on {\sl HST} images for 126 knots
covering a nine month period which appear to be among the least decelerated
ejecta.  Based on these studies, we will adopt an explosion date of 1680, thus
making the remnant's current age to be 329 yr. 

This age yields a free expansion proper motion of $0 \farcs 465$ yr$^{-1}$, or,
assuming a distance of 3.4 kpc, a free expansion velocity of $\approx$ 7500 km
s$^{-1}$. We can thus calculate the deceleration parameter of the blastwave as
$m$ = (4900 km s$^{-1}$/7500 km s$^{-1}$) $\approx$ 0.65, or
equivalently, using Gotthelf et
al.'s angular remnant size in 2000, 
$0 \farcs 30$ yr$^{-1}$/(153$\arcsec$ $\pm$ 12$\arcsec$ / 320 yr)
$\approx$ $0.58 - 0.68$. 

\subsection{Cas A Expansion Models}

Our measurements of Cas A's forward shock proper motion and estimated
deceleration parameter can be used to model the SNR's evolution.  In
ejecta--dominated remnants, the deceleration parameter is related to the
self-similar evolution by $m$ = $(n-3)/(n-s)$
\citep{chevalier82,truelove99,laming03}, where $n$ is the power-law index for
the ejecta density profile ($\rho_{ej} \propto r^{-n}$) and $s$ is the
power-law index for the ambient medium density profile ($\rho_{amb} \propto
r^{-s}$).  Generally, $s=0$ corresponds to a constant density ambient medium,
while $s=2$ corresponds to an ambient medium shaped by a circumstellar wind.
For the progenitors of core-collapse SNe, such as Cas A, $s=2$. 

For remnants in the adiabatic (Sedov-Taylor) stage of expansion, the deceleration
parameter $m=0.67$.  Many young remnants, such as Tycho, Kepler, SN~1006, and
Cas A, are believed to be currently transitioning between the ejecta--dominated
and Sedov stage. However, our calculated deceleration parameter of 0.65 is less
than that expected for Sedov-type expansion, and corresponds to an ejecta
power-law index of 4.85.

However, \citet{laming03} estimated a much higher ejecta density profile for Cas A.
Using a Lagrangian hydrodynamics model coupled to a non-equilibrium ionization
code, they self-consistently modeled the density profile of Cas A's expanding
ejecta and found found that the ejecta density is well described by a power-law
of index $n=7-9$. This corresponds to a deceleration parameter of $m=0.8-0.86$,
considerably larger than our derived deceleration parameter of 0.65.

\citet{truelove99} point out that for models for SNR evolution in which $3 < n
< 5$, the bulk of the mass is concentrated at lower velocities, while the bulk
of the energy is concentrated at higher velocities.  Furthermore, the timescale
by which a SNR enters the Sedov-Taylor phase of its evolution is set by the
time that the reverse shock takes to travel through ejecta containing the bulk
of the energy. Thus, in models with mass-poor and energy rich envelopes, this
transition time can be very short. \citet{laming03} suggest that Cas A is
currently transitioning from the ejecta-dominated to the Sedov-Taylor phase, so
a power-law index as low as our estimated value of 4.85 seems unlikely.

In order to understand this discrepancy, we have tried to model Cas A's
expansion.  At an assumed distance of 3.4 kpc and a 320 yr age in 2000, Cas A's
average forward shock radius of 153$\arcsec$ translates to 2.5 pc in radius and
an average reverse shock radius $95'' \pm 10''$ corresponding to $1.6 \pm 0.2$
pc.

We adopted \citet{laming03} estimated explosion energy of $2 \times$
10$^{51}$ erg and ejecta mass of 2 M$_{\sun}$, assume that the SNR is
expanding into a red giant wind \citep{Krause08}, and choose $v_{wind}$
$\approx$ 10 km s$^{-1}$ and $\dot{M}$ $\approx$ 2 $\times$ 10$^{-5}$
M$_{\sun}$ yr$^{-1}$. The results of these adopted values, summarized in Model $1$ in
Table~\ref{tab:snrmodels}, show that our estimated ejecta power-law index of
4.85 does not reproduce the Cas A's measured parameters, producing a forward
shock radius of 2.93 pc and velocity of 6300 km s$^{-1}$ instead of the 2.5 pc
and $\simeq$5000 km s$^{-1}$ values actually observed assuming a distance of
3.4 kpc. 

Given that our initial derived ejecta power-law index does not agree with that 
derived from
spectral fits to the SNR ejecta, we explored models with ejecta profiles
consistent with Laming \& Hwang's fits (Models $2-7$ in
Table~\ref{tab:snrmodels}). We note that a similar set of parameters were also
chosen by \citet{schure08} in the context of Cas A's jet evolution in a
Wolf-Rayet bubble, although their models do not consistently match both the
observed blastwave radius and velocity either (see their Table~1).

As shown in Table~\ref{tab:snrmodels}, while our Models $2-7$ may be
appropriate for the evolution of the SNR ejecta and the jet, they overestimate
the forward shock velocity regardless of choice of the power-law index of the
ejecta or progenitor wind structure.  These models also do not fit the measured
expansion of Cas A, producing deceleration parameters of $m > 0.7$ and shock
velocities $v_{\mathrm{shock}} > 5500$ km s$^{-1}$. 

\subsection{Cosmic Ray Acceleration at the Forward Shock} 

As there is a great deal of evidence suggesting that shocks in SNRs
are efficient generators of cosmic rays \citep[e.g.,][]{warren05}, 
we then explored the inclusion of
cosmic ray modification of the forward shock as a possible solution to these
poor model fits.
A signature of shock generated cosmic rays are nonthermal 
X-rays generated by synchrotron radiation due to shock-accelerated TeV
electrons. High energy photons at GeV--Tev energies, either inverse Compton
radiation from electrons or pion-decay emission from ions, have
been detected from some supernova remnants including Cas A 
by HEGRA \citep{aharonian01} and MAGIC \citep{albert07}.

In the production of cosmic rays, energy is removed from the SNR shock via
particle acceleration.  In doing so, the shock slows and the post-shock gas
becomes more compressed.  We therefore also modeled Cas A under this
assumption.  

The inclusion of efficient acceleration at the forward shock should not alter
the dynamics of the ejecta, and thus these models can be consistent with
\citet{laming03}.  We also chose to only model shock acceleration at the
forward shock.  Although there have been suggestions that the bulk of the
particle acceleration in Cas A might be occurring at the reverse shock
\citep{uchiyama08,helder08}, the degree to which particle acceleration at the
reverse shock is efficient remains an open question (see below).  

Starting with the parameter space explored by \citet{laming03}, we modeled Cas~A
assuming that some fraction of the explosion energy has gone into
accelerating cosmic rays. These models were set up as in \citet{ellison07}
where the nonlinear particle acceleration is tuned by an injection parameter
which determines the fraction of thermal particles that are injected into the
acceleration process thus determining how much of the energy of the SNR
goes into cosmic rays.  These models are listed as Models 8--15 in
Table~\ref{tab:snrmodels}. The particle injection is sensitive to parameters
such as the shock velocity and ambient density, so choosing a fixed injection
while varying the environmental parameters will naturally lead to varying
acceleration efficiencies, as seen in Models 14--15.

As expected and shown in Table~\ref{tab:snrmodels}, increasingly efficient
particle acceleration leads to lower shock velocities and smaller forward shock
radii, leading to smaller modeled expansion parameters. In Models 8--13, we
attempted to tune the acceleration efficiency so as to match the measured
forward shock expansion velocity and forward and reverse shock radii. We fixed
the ejecta density distribution as well as the explosion and pre-supernova wind
parameters, and in Models 11--13, we fixed the acceleration efficiency but
varied the ejecta power-law index.

We found that Models 11--13, with power-law indices of $n=7-9$, a wind velocity
of $v_{\mathrm{wind}}$ of 10 km s$^{-1}$, and a progenitor pre-SN mass loss
rate of $\dot{M}$ $\approx$ 2 $\times$ 10$^{-5}$ M$_{\sun}$ yr$^{-1}$ provide
a good fit to our observations, where $\gtrsim$ 30\% of the SN explosion energy
is lost in particle acceleration. This acceleration efficiency results in a
modeled forward shock velocity of $\sim$ 5000 km s$^{-1}$, forward and reverse
shock radii of $2.46$ pc and $1.67$ pc, and a deceleration parameter of $m$ = 0.66.
These values agree well with our measured deceleration parameter of
0.65 and measured blastwave velocity of 4900 km s$^{-1}$, while also being
consistent with the spectral fits of \citet{laming03} and the measured forward
and reverse shock radii of $2.5$ pc and $1.6$ pc \citep{gotthelf01}. 
We found that
varying the ejecta power-law index, has only a small effect on the final
parameters, seen as a difference in the blastwave velocity in Models 11--13.

We also tried varying the pre-supernova wind parameters in Models 14--15 to
match those of \citet{schure08}. While these models result in similar
deceleration parameters and forward shock radii to Models 11--13, they
significantly overestimate the forward shock velocity.

Finally, in order to see if our results could be fit by models that do not
include the effects of diffusive shock acceleration, we also explored a wider
parameter space in both the ejecta mass and explosion energy. These are listed
as Models $16 - 23$ in Table~\ref{tab:snrmodels}, where in Models $16 - 21$ we
varied the explosion energy and ejecta mass between 1.0--2.0 $\times$ 10$^{51}$
erg and 1.0--2.0 M$_{\sun}$.  In Models $22 - 23$, we only varied the explosion
energy, while fixing the other parameters as in Model 2.

As seen in Table~\ref{tab:snrmodels}, varying the explosion energy and ejecta
mass does not allow for a simultaneous fit of both the forward shock radius and
velocity. For example, in Model 17 we find a suitable fit to the the forward
shock radius, but the reverse shock radius is too small and the forward shock
velocity is too high. Conversely, in Models 16 and 19 the forward shock
velocity is well fit, but the forward shock radius is too small.  While it is
conceivable that one could design a model which can simultaneously fit the
forward shock radius and blastwave velocity, such a model might not be 
consistent with other parameters derived from spectral fits to the ejecta.

Although our modeling results suggest significant cosmic ray production at the forward
shock, it is uncertain whether efficient particle acceleration might also be
occurring at the reverse shock as well \citep{ellison05}.  If efficient shock
acceleration were occurring at the reverse shock, other effects of this
acceleration would be directly observable, both in the dynamics of the reverse
shock and in the emitted thermal spectrum \citep{ellison05,ellison07}.  Much
like in shock acceleration at the forward shock, the process removes energy
from the shock and softens the equation of state. If particle acceleration were
efficient, we would expect to observe the reverse shock to be closer to the
contact discontinuity (much like the forward shock is close to the contact
discontinuity in Tycho's SNR; \citealt{warren05}). The fact that our cosmic
ray models appear to predict with good accuracy the location of the reverse
shock suggests that efficient acceleration may not be present at significant
levels at the reverse shock. Furthermore, 
the presence of of Fe-K emission at the reverse shock, seen in
equivalent width maps \citep{hwang00} suggests a high shock (and electron)
temperature at the reverse shock, suggesting that the reverse shock has not
lost much energy to cosmic ray acceleration.

\subsection{Brightness Variations of Nonthermal X-ray 
Filaments and their Origin} 

Rapid changes in the brightness of thin, nonthermal filaments in the interior
of Cas A have been noted previously via comparisons of the 2000--2004
observations \citep{patnaude07,uchiyama08}.  A comparison of all four epoch
{\sl Chandra} ACIS images, covering nearly an eight year time span, highlights
and clarifies many of these changes in filament brightness and position. This
is most readily seen in an on-line movie where we show the evolution of Cas A's
X-ray emission between 2000 and 2007, of which Figure~\ref{fig:casa07} is but
one frame. 

\begin{figure}
\plotone{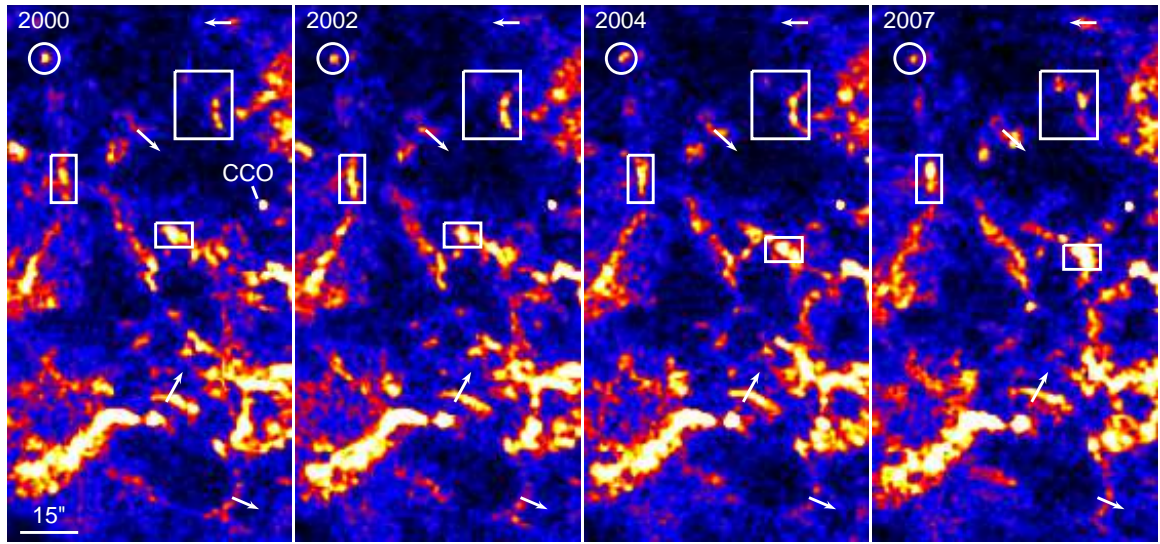}
\caption{The east-central region of Cas A in the 4.2--6.0 keV band. 
The four frames
show the central region between 2000 and 2007. Boxes and the circle mark
knots and filaments which show brightness variations along the filament,
while arrows mark the location and direction of thin filaments which 
show proper motions between 2000 and 2007. The central compact object (CCO)
is labeled for reference.}
\label{fig:center}
\end{figure}

A close-up view of many of the changes exhibited by interior nonthermal
emission features is presented in Figure~\ref{fig:center}, where we show the
east-central region of Cas A in each epoch in the 4.2--6.0 keV band. In these
images the remnant's global structure of continuum emission appears not unlike
that seen in the radio; that is, the emission is characterized by thin,
web-like and highly filamentary structures which often enclose patchy, faint
diffuse emission. 

A comparison of the four frames in Figure~\ref{fig:center} reveals several
regions where the continuum emission dramatically brightens or fades between
Jan 2000 and Dec 2007.  Sections of some nonthermal filaments change so
substantially between images that they resemble apparent rapid proper motions
($\simeq$ $0 \farcs 2 - 0 \farcs 3$ yr$^{-1}$) that are, in some places,
directed inward toward the remnant center or at some random, often non-radial
direction.  In addition, apparent sequential brightening of small sections 
of some filaments can give the appearance of motion along the filament. 

\begin{figure}
\plottwo{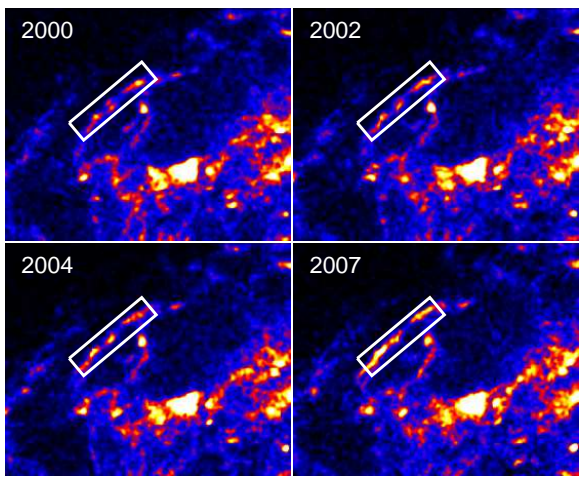}{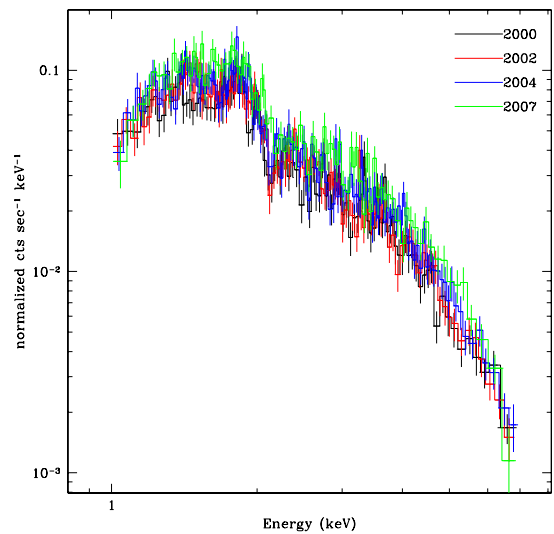}
\caption{{\it left}: Exposure corrected 4.2--6.0 keV images of a
bright nonthermal filament (enclosed in the white box) in
the northeast corner of Cas A. {\it right}: Spectral fits the the spectrum
from this filament. The fit results are listed in Table~\ref{tab:fits}.}
\label{fig:ne}
\end{figure}

Whereas the bulk of the changes in the remnant's nonthermal emission appear to
come from knots and filaments which lie inside or projected onto the interior
of the SNR, a few outer forward shock front filaments can also show similar
changes in brightness.  One filament associated with the forward shock, shown
in Figure~\ref{fig:ne}, shows evidence for substantial brightening between 2000
and 2007, with non-radial sequential changes seen along its length.  This
filament had previously been identified by \citet{stage06} as a potential site
for efficient shock acceleration, and our new observations confirm that the
filament exhibits behavior consistent with the changes seen in the interior
filaments.

\citet{uchiyama08} argue that emission flaring of nonthermal filaments is
evidence for electron acceleration while a decrease in flux corresponds to
synchrotron cooling.  Using the {\sl Chandra} ACIS 2000--2004 data, they found
such emission flaring and fading was most apparent in interior filaments,
leading them to conclude that such particle acceleration and synchrotron
cooling was more likely to be occurring at the reverse shock, a 
conclusion supported by the deprojected continuum images of Cas A
presented by \citet{helder08}.

However, the addition of the new Dec 2007 observations which increases the
timespan from 4 to nearly 8 years shows clear evidence for brightness variations of
outer nonthermal filaments associated with the forward blastwave.  As shown in
Figure~\ref{fig:ne} and listed in Table~\ref{tab:fits}, the northeast filament
brightens substantially between Jan 2004 and Dec 2007.  Hence, rapid electron
acceleration would appear to be occurring in some forward shock filaments as
well.  

In cases of increasing X-ray flux, the acceleration time of an X-ray
emitting electron is given by $t_{acc} \sim 9 \eta B^{-3/2}_{\rm mG}
\varepsilon^{1/2}_{\rm keV} V^{-2}_{1000}$ yr, where $\eta \geq 1$ is the
electron gyro-factor, $V_{1000}$ is the shock velocity in units of 1000 km
s$^{-1}$, and $\varepsilon_{\rm keV}$ is the mean photon energy ($\approx$ 1
keV).  As listed in Table~\ref{tab:pm}, the mean proper motion of this filament
is $\sim$ $0 \farcs 30$ yr$^{-1}$, which at a distance of 3.4 kpc corresponds
to $V_{1000}$ = 4.9. 

\citet{uchiyama08} have suggested that such brightness changes in the remnant's
interior nonthermal emission filaments originate at the remnant's reverse shock
(due to their projected interior position), a notion first suggested by
\citet{bleeker01} based on hardness ratios for interior and outer shock
filaments as measured from {\sl XMM--Newton} images. Support for the
interpretation that the exterior and interior nonthermal emission filaments
arise from different sources is the lack of radio emission associated with the
exterior X-ray forward shock filaments, in contrast to the fair correlation
that exists between interior radio and X-ray filaments \citep{delaney04b}.
\citet{helder08} have also concluded that the interior nonthermal filaments
originate from the reverse shock and not the forward shock.

On the other hand, \citet{delaney04a} and \citet{delaney04b} have argued that
interior nonthermal filaments may merely be forward shock filaments seen in
projection against the face of Cas A. In this view, interior filamentary and
web-like structures arise as the forward shock interacts with a lumpy,
inhomogeneous CSM, with the observed brightness variations arising from line of
sight tangencies of the shock front as it progresses through and around small
CSM clouds and density variations.  

We note that a correlation between global X-ray and radio filaments is not
expected, thus undermining the meaning of any correlation of nonthermal radio
and X-ray emitting features. Both \citet{CC05} and \citet{ellison05b} showed
that in the remnants of core-collapse SNe interacting with a stellar wind,
the non-thermal X-ray emission is strongly peaked at the shock front while
radio emission will peak at the contact discontinuity. This can be seen in
Figure~8 of \citet{ellison05b} where the peak radio emissivity occurs well
inside of the X-ray (see Figure~4 of \citet{CC05} for another example). 

\begin{figure}
\includegraphics[width=3.5in, bb=221 114 584 708]{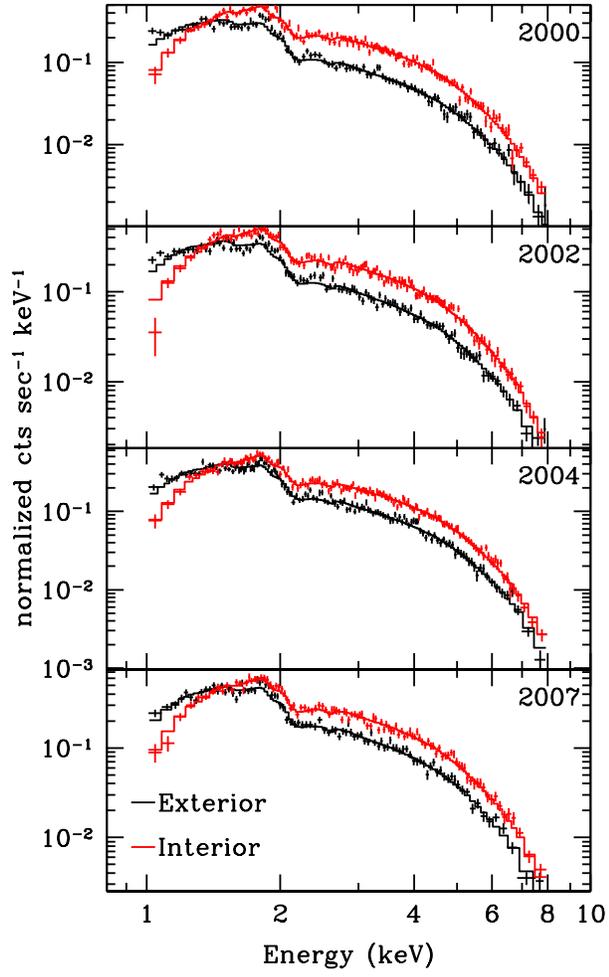}
\caption{Extracted spectra for forward shock and interior nonthermal 
filaments.}
\label{fig:fits}
\end{figure}

To investigate the question of whether the nonthermal filaments
projected in the interior of Cas A are associated with the reverse shock or the
forward shock, we extracted spectra for six exterior forward shock filaments
(including the NE filament marked in Fig.\ \ref{fig:ne}) and 23 interior
projected nonthermal filaments from our Dec 2007 observations using the CIAO
tool {\tt specextract}.  We also extracted spectra for these same filaments
from the 2000, 2002, and 2004 data. These data were then fit with absorbed
power-laws. The results from these spectral fits for both exterior and interior
filaments are listed in Table~\ref{tab:fits} and plotted in
Figure~\ref{fig:fits}. 

Aside from obvious normalizations and differences in the absorbing column
affecting the flux at lower energies, the spectra for exterior and interior
nonthermal filaments are qualitatively quite similar (Fig.\ \ref{fig:fits}).
As shown in Table~\ref{tab:fits}, while the fitted spectral indices hardly
differ, interior filaments do appear to be marginally harder consistent
with the conclusion of \citet{bleeker01}. 

\subsection{Magnetic Field Strength}

Lastly, we turn to the question of magnetic field strength in the filaments.
As noted above, the northeast filament shows evidence for
brightness changes over a nearly eight year timespan.  If we adopt an
acceleration time $t_{acc}$ $\sim$ 2--8 yr, then this corresponds to a magnetic
field strength of $B_{\rm mG}$ $\sim$ 0.1--0.3, with the lower limit
corresponding to the upper limit on the acceleration time. Our results are
consistent with magnetic field strengths derived from previous observations
\citep{longair94,wright99,vink03,atoyan00,berezhko04} as well as the 
recent results of \citet{uchiyama08}.

Recently, \citet{bykov08} simulated the effects of magnetic field turbulence on
the observed synchrotron emission in young SNRs. They showed that the structure
and evolution of small clumps ($\sim$ 10$^{14}$ -- 10$^{16}$ cm) can change on
timescales $\sim$ 1 year.  The angular size of the knots and filaments seen in
Figure~\ref{fig:center} is $\sim$ 5$\arcsec$ which corresponds to $\sim$ 2.5
$\times$ 10$^{17}$ cm at Cas A's estimated a distance of 3.4 kpc. Significant
flux variations on this spatial scale are seen to occur over the time period of
$\sim$ 4 yr, meaning that that yearly changes could occur over $\sim$ 6
$\times$ 10$^{16}$ cm. 

\citet{bykov08} argue that intensity variations on such spatial scales are
consistent with localized regions of high magnetic field ($\gtrsim$ 0.1 mG),
brought about by turbulence behind the shock. Furthermore, they point out
that the integrated line of sight emissivity of these knots and filaments is
what allows them to stand out against background emission. In \citet{bykov08},
the shock is propagating perpendicular to the line of sight, but similar
results are expected to be visible in face-on-shocks \citep{bykov09}, 
consistent with our observations of flux changes seen in both the 
exterior and face--on filaments.

\section{Conclusions}

We have presented new {\it Chandra} ACIS observations of Cas A which were taken
in late 2007. These new observations, when combined with previous {\it Chandra}
data, allow us to constrain the velocity of the forward shock to be about 4900
km s$^{-1}$.  

Combined with results from previous analyses of Cas A's X-ray emission
\citep{laming03,gotthelf01}, we present several models for the evolution of Cas
A and find that it's expansion can be well modeled by an $n= 7 - 9$ ejecta profile
running into a circumstellar wind.  We also find that the position of the
reverse shock in this model is consistent with that measured by
\citet{gotthelf01}. However, in order to match the radius of the forward shock,
we found that we must assume that the forward shock is efficiently accelerating
cosmic rays.

Rapid changes in Cas A's synchrotron emission are seen for interior and
exterior projected filaments, with both showing similar nonthermal spectra as
well as inferred magnetic field strengths.  Based on this and the simulations
presented by \citet{bykov08}, it is currently not clear whether the interior
filaments are in fact located at the reverse shock as recently argued by
\citet{uchiyama08} and \citet{helder08}.  

Instead, we propose that the interior
filaments might be forward shocks seen in projection \citep{delaney04b}.  In
that case, the observed brightness variations might arise from wrinkles in
front-facing, forward shock as it moves through an inhomogeneous, local
circumstellar medium.

Although we cannot rule out the possibility that interior nonthermal filaments
are associated with the reverse shock, the combination of similar spectra,
flaring timescale, and our fits to the remnant's dynamics are suggestive that
the observed synchrotron flaring for interior filaments arises from forward
shock filaments seen in projection toward Cas A's interior rather than at the
reverse shock as recently suggested.  At the least, our new X-ray data of Cas A
shows that rapid brightness variations like those seen for interior nonthermal
filaments can also be exhibited by some outer, nonthermal forward shock
filaments.

\acknowledgements

We thank Don Ellison, Stephen Reynolds, and Martin Laming for many useful
discussions during the preparation of this paper. We also wish to thank the
anonymous referee whose many suggestions and corrections significantly 
improved the
paper.  This work was supported by NASA grant GO8-9065A which is administered
by the CXC at SAO. D.~J.~P. acknowledges support from NASA contract NAS8-39073.

{\it Facilities:} \facility{CXO (ACIS)}.

\clearpage

\begin{deluxetable}{rccccc}
\tablecolumns{6}
\tablewidth{0pc}
\tablecaption{Forward Shock Filament Proper Motions}
\tablehead{
\colhead{Region\tablenotemark{a}} & \colhead{DeLaney \& Rudnick (2003)\tablenotemark{b}} & \multicolumn{2}{c}{Cross Correlation} & \multicolumn{2}{c}{Profile Fits} \\
\colhead{} & \colhead{} & \colhead{2000--2002} & \colhead{2000--2007} &
\colhead{2000--2002} & \colhead{2000--2007} \\
\colhead{} & \multicolumn{5}{c}{$\arcsec$ yr$^{-1}$}}
\startdata
Southeast & $0 \farcs 38 \pm 0 \farcs 03$ & $0 \farcs 31 \pm 0 \farcs 04$ & $0 \farcs 31 \pm 0 \farcs 02$ & $0 \farcs 33 \pm 0 \farcs 03$ & $0 \farcs 32 \pm 0 \farcs 04$ \\
East      & $0 \farcs 38 \pm 0 \farcs 03$ & $0 \farcs 30 \pm 0 \farcs 03$ & $0 \farcs 31 \pm 0 \farcs 02$ & $0 \farcs 30 \pm 0 \farcs 02$ & $0 \farcs 32 \pm 0 \farcs 02$ \\
Northeast & $0 \farcs 41 \pm 0 \farcs 02$ & $0 \farcs 34 \pm 0 \farcs 04$ & $0 \farcs 31 \pm 0 \farcs 04$ & $0 \farcs 30 \pm 0 \farcs 04$ & $0 \farcs 30 \pm 0 \farcs 03$ \\
North     & $0 \farcs 28 \pm 0 \farcs 01$ & $0 \farcs 29 \pm 0 \farcs 03$ & $0 \farcs 26 \pm 0 \farcs 02$ & $0 \farcs 28 \pm 0 \farcs 03$ & $0 \farcs 27 \pm 0 \farcs 02$ \\
Northwest & \nodata                       & $0 \farcs 25 \pm 0 \farcs 06$ & $0 \farcs 27 \pm 0 \farcs 04$ & $0 \farcs 28 \pm 0 \farcs 04$ & $0 \farcs 28 \pm 0 \farcs 03$ \\
South     & \nodata                       & $0 \farcs 31 \pm 0 \farcs 02$ & $0 \farcs 32 \pm 0 \farcs 04$ & $0 \farcs 32 \pm 0 \farcs 05$ & $0 \farcs 34 \pm 0 \farcs 03$\\
\enddata
\tablenotetext{a}{The southeast, east, northeast, and north regions 
correspond to Regions 26, 29, 2, and 14 respectively in \citet{delaney03}.}
\tablenotetext{b}{Filament velocities from Table~2 of \citet{delaney03}
and converted to proper motions assuming a distance of 3.4 kpc.}
\label{tab:pm}
\end{deluxetable}

\begin{deluxetable}{rcccccccccc}
\tablecolumns{11}
\tablewidth{0pc}
\tablecaption{Cas A Evolutionary Models }
\tablehead{
\colhead{} & \colhead{M$_{ej}$} & \colhead{E$_{51}$} & \colhead{} &
\colhead{$v_{wind}$} &\colhead{$\dot{M}_{-5}$} & \colhead{E(CR)/E(SN)} &
\colhead{R$_{FS}$} & \colhead{R$_{RS}$} & \colhead{V$_{shock}$} &
\colhead{} \\
\colhead{Model} & \colhead{M$_{\sun}$} & \colhead{10$^{51}$ erg} & 
\colhead{$n$} &
\colhead{km s$^{-1}$} & \colhead{10$^{-5}$ M$_{\sun}$ yr$^{-1}$} & 
\colhead{\%} &
\colhead{pc} & \colhead{pc} & \colhead{km s$^{-1}$} & \colhead{$m$}
}
\startdata
1 & 2 & 2 & 4.85 & 10 & 2 & 0 & 2.93 & 1.61 & 6300 & 0.70 \\
2 & 2.5 & 2 & 9 & 5 & 1.5 & 0 & 2.44 & 1.42 & 5500 & 0.73 \\
3 & 2 & 2 & 9 & 10 & 2 & 0 & 2.79 & 1.65 & 6376 & 0.74 \\
4 & 2 & 2 & 8 & 10 & 2 & 0 & 2.78 & 1.67 & 6379 & 0.74 \\
5 & 2 & 2 & 7 & 10 & 2 & 0 & 2.78 & 1.67 & 6390 & 0.74 \\
6 & 2 & 2 & 6 & 10 & 2 & 0 & 2.78 & 1.73 & 6352 & 0.74 \\
7 & 2 & 2 & 9 & 5 & 1.5 & 0 & 2.56 & 1.42 & 5679 & 0.72 \\
8 & 2 & 2 & 9 & 10 & 2 & 7 & 2.73 & 1.67 & 6178 & 0.73 \\
9 & 2 & 2 & 9 & 10 & 2 & 50 & 2.22 & 1.67 & 4613 & 0.67 \\
10 & 2 & 2 & 9 & 10 & 2 & 17 & 2.64 & 1.67 & 5826 & 0.72 \\
11 & 2 & 2 & 9 & 10 & 2 & 34 & 2.46 & 1.67 & 5021 & 0.66 \\
12 & 2 & 2 & 8 & 10 & 2 & 34 & 2.46 & 1.67 & 5023 & 0.66 \\
13 & 2 & 2 & 7 & 10 & 2 & 34 & 2.46 & 1.67 & 5033 & 0.66 \\
14 & 2 & 2 & 9 & 10 & 1.5 & 27 & 2.68 & 1.85 & 5594 & 0.68 \\
15 & 2 & 2 & 9 & 5 & 1.5 & 27 & 2.68 & 1.85 & 5594 & 0.68 \\
16 & 2 & 1 & 9 & 10 & 2 & 0 & 2.14 & 1.36 & 5010 & 0.77 \\
17 & 2 & 1.5 & 9 & 10 & 2 & 0 & 2.50 & 1.54 & 5768 & 0.76 \\
18 & 1 & 1 & 9 & 10 & 2 & 0 & 2.40 & 1.17 & 5215 & 0.71 \\
19 & 1.5 & 1 & 9 & 10 & 2 & 0 & 2.26 & 1.30 & 5120 & 0.74 \\
20 & 1 & 1.5 & 9 & 10 & 2 & 0 & 2.77 & 1.30 & 5989 & 0.71 \\
21 & 1.5 & 1.5 & 9 & 10 & 2 & 0 & 2.63 & 1.48 & 5861 & 0.73 \\
22 & 2.5 & 1 & 9 & 4.7 & 1.5 & 0 & 1.88 & 1.17 & 4348 & 0.76 \\
23 & 2.5 & 1.5 & 9 & 4.7 & 1.5 & 0 & 2.20 & 1.30 & 4994 & 0.74 \\
\enddata
\label{tab:snrmodels}
\end{deluxetable}

\begin{deluxetable}{rcccccc}
\tablecolumns{7}
\tablewidth{0pc}
\tablecaption{Nonthermal Filaments Spectral Fits}
\tablehead{
\colhead{} & \multicolumn{2}{c}{Exterior Filaments\tablenotemark{a}} 
& \multicolumn{2}{c}{Interior Filaments\tablenotemark{b}} 
& \multicolumn{2}{c}{Northeast Filament\tablenotemark{c}} \\
\colhead{Epoch} & \colhead{$\Gamma$} & \colhead{1 keV Flux\tablenotemark{d}} 
& \colhead{$\Gamma$} & \colhead{1 keV Flux\tablenotemark{d}}
& \colhead{$\Gamma$} & \colhead{1 keV Flux\tablenotemark{d}}}
\startdata
2000 & 2.27$^{+0.08}_{-0.08}$ & 2.81$^{+0.31}_{-0.28}$ & 2.41$^{+0.07}_{-0.07}$ & 8.10$^{+0.81}_{-0.73}$ & 2.28$^{+0.09}_{-0.08}$ & 0.78$^{+0.08}_{-0.08}$\\
2002 & 2.27$^{+0.07}_{-0.07}$ & 3.31$^{+0.32}_{-0.29}$ & 2.41$^{+0.06}_{-0.06}$ & 8.59$^{+0.81}_{-0.73}$ & 2.39$^{+0.08}_{-0.08}$ & 0.96$^{+0.10}_{-0.09}$\\
2004 & 2.31$^{+0.06}_{-0.06}$ & 4.03$^{+0.36}_{-0.33}$ & 2.41$^{+0.06}_{-0.06}$ & 8.94$^{+0.83}_{-0.75}$ & 2.31$^{+0.08}_{-0.08}$ & 1.01$^{+0.10}_{-0.09}$\\
2007 & 2.35$^{+0.08}_{-0.08}$ & 5.15$^{+0.57}_{-0.51}$ & 2.36$^{+0.07}_{-0.07}$ & 10.1$^{+0.11}_{-0.12}$ & 2.29$^{+0.09}_{-0.09}$ & 1.18$^{+0.13}_{-0.12}$\\
\enddata
\tablenotetext{a}{Galactic $N_H$ fit at 1.15 $\times$ 10$^{22}$ cm$^{-2}$.}
\tablenotetext{b}{Galactic $N_H$ fit at 2.01 $\times$ 10$^{22}$ cm$^{-2}$.}
\tablenotetext{c}{Galactic $N_H$ fit at 1.22 $\times$ 10$^{22}$ cm$^{-2}$.}
\tablenotetext{d}{in units of 10$^{-3}$ photons keV$^{-1}$ cm$^{-2}$ s$^{-1}$}
\label{tab:fits}
\end{deluxetable}

\clearpage

\end{document}